\documentclass[graybox]{SNmult}

\usepackage{type1cm}        %
\usepackage{makeidx}         %
\usepackage{graphicx}        %
\usepackage{multicol}        %
\usepackage[bottom]{footmisc}%

\usepackage{newtxtext}       %
\usepackage[varvw]{newtxmath}       %

\makeindex             %

\usepackage{subcaption}
\usepackage{stmaryrd}
\usepackage{mathtools}
\usepackage{siunitx}
\DeclareMathOperator{\seibelCurlSD}{\text{curl}_\text{2D}}
\DeclareMathOperator{\seibelCurl}{\textbf{curl}}
\DeclareMathOperator{\seibelDiv}{div}
\DeclareMathOperator{\seibelGrad}{\textbf{grad}}
\newcommand{\seibelOmG}{{\Omega\setminus\Gamma}}
\newcommand{\seibelJump}[1]{\big\llbracket #1 \big\rrbracket}

\begin{document}
\title*{On the Application of Hybrid Mixed Domain Decomposition Methods to Permanent Magnet Synchronous Machines}
\titlerunning{HMDD for PMSM}
\author{Timon Seibel\orcidID{0009-0007-2570-0248} and\\ Sebastian Schöps\orcidID{0000-0001-9150-0219} and\\ Kersten Schmidt\orcidID{0000-0001-7729-6960}}
\authorrunning{Timon Seibel et al.}
\institute{Timon Seibel \at Technical University of Darmstadt, \email{timon.seibel@tu-darmstadt.de}
\and Sebastian Schöps \at Technical University of Darmstadt, \email{sebastian.schoeps@tu-darmstadt.de}
\and Kersten Schmidt \at Technical University of Darmstadt, \email{kschmidt@mathematik.tu-darmstadt.de}}
\maketitle
\abstract*{In this work, we study the application of a \emph{hybrid mixed domain decomposition} (HMDD) method for the rotor-stator coupling of a \emph{permanent magnet synchronous machine}. For this, we derive a variational formulation on the electric machine inspired by \emph{hybridized discontinuous Galerkin} methods using a mixed magnetostatics problem, an affine material law and boundary conditions respecting the symmetry of the motor. We are then able to locate the resulting finite element method within the HMDD framework. This enables us naturally to transfer the well-posedness results and error estimates for the HMDD method to the finite element method considered in this work. Lastly, as a proof of concept, we consider an academic example and compare the resulting magnetic flux density and potential lines to their counterparts obtained by a well-established in-house code using \emph{iso-geometric analysis}.
}

\abstract{In this work, we study the application of a \emph{hybrid mixed domain decomposition} (HMDD) method for the rotor-stator coupling of a \emph{permanent magnet synchronous machine}. For this, we derive a variational formulation on the electric machine inspired by \emph{hybridized discontinuous Galerkin} methods using a mixed magnetostatics problem, an affine material law and boundary conditions respecting the symmetry of the motor. We are then able to locate the resulting finite element method within the HMDD framework. This enables us naturally to transfer the well-posedness results and error estimates for the HMDD method to the finite element method considered in this work. Lastly, as a proof of concept, we consider an academic example and compare the resulting magnetic flux density and potential lines to their counterparts obtained by a well-established in-house code using \emph{iso-geometric analysis}.
}

\section{Introduction}\label{Seibel:sec:introduction}
\begin{figure}[h]
	\begin{subfigure}{.45\textwidth}
		\includegraphics[height=4.5cm]{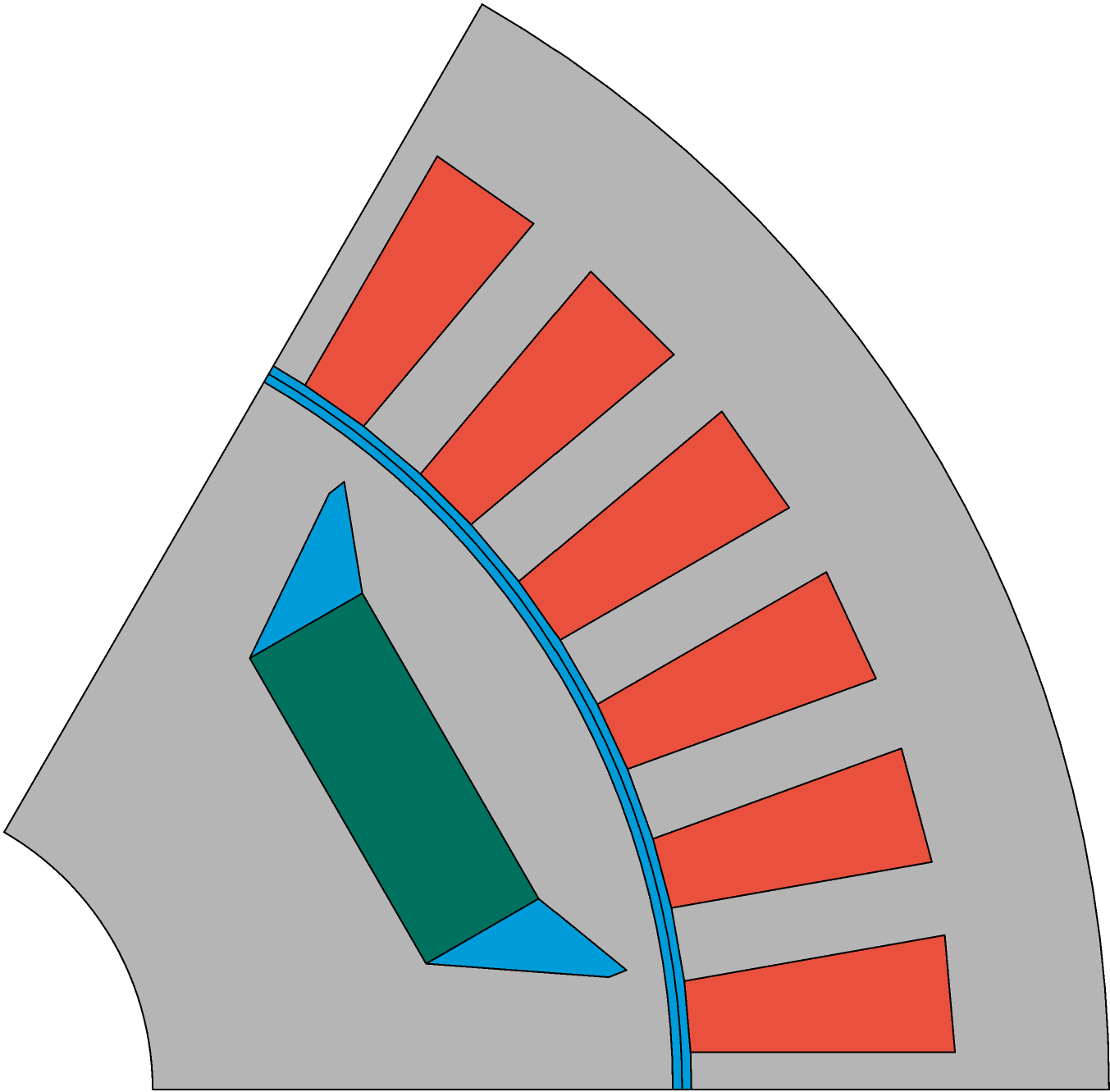}
		\subcaption{Sketch of a simplified Permanent Magnet Synchronous Machine. Inspired by Fig.~4 of \cite{Seibel:Wiesheu_2024aa}. The permanent magnet is shown in green ($\mu_r = \num{1.05}$), the coils in red, the air gaps in blue (both $\mu_r = \num{1}$) and iron rotor and stator in gray ($\mu_r = \num{1000}$).}
		\label{Seibel:fig:PMSM}
	\end{subfigure}
	\qquad
	\begin{subfigure}{.45\textwidth}
		\includegraphics[height=4.5cm]{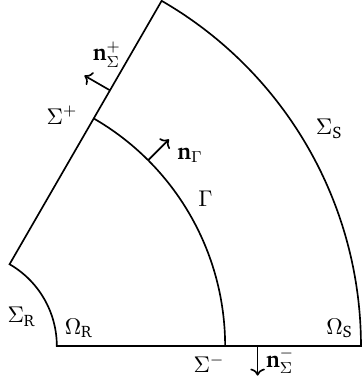}
		\caption{Sketch of the computational domain with the rotor domain $\Omega_\text{R}$ and the stator domain $\Omega_\text{S}$ -- both incorporating half of the air gap --, the interface $\Gamma$ between $\Omega_\text{R}$ and $\Omega_\text{S}$ and the outer boundaries $\Sigma^-$, $\Sigma^+$, $\Sigma_\text{R}$ and $\Sigma_{S}$.}
		\label{Seibel:fig:CompDom}
	\end{subfigure}
	\caption{Simplified sketch of a PMSM (left) and the computational domain (right), resp.}
	\label{Seibel:fig:PMSMandCD}
\end{figure}

Electric machines are widely used in modern applications and their efficient numerical simulation remains an important topic in computational engineering. In this work, we study \emph{permanent magnet synchronous machines} (PMSM) and exploit their periodic structure by restricting the analysis to a single pole (cf. \cite{Seibel:Salon_1995aa}). Due to their layered, rotationally symmetric design and the clear separation into rotor, stator and air gaps, electric motors are naturally well-suited to domain decomposition approaches such as \emph{hybrid mixed domain decomposition} (HMDD) \cite{Seibel:Schmidt_2026aa}, as is evident from Fig.~\ref{Seibel:fig:PMSM}.

The main contribution of this work is the application of HMDD to a magnetostatic rotor-stator coupling in a geometry-respecting high-order finite element setting. In addition, we transfer the well-posedness results and error estimates from \cite{Seibel:Schmidt_2026aa} and present an academic proof-of-concept example.

\section{Model Problem}
We consider a 2D mixed formulation of the magnetostatic Maxwell equations: 
\begin{eqnarray}
	\seibelCurl \vec{H} = \vec{J},\qquad\seibelDiv\vec{B} = 0,\label{Seibel:eq:MXWdivB0}
\end{eqnarray}
where $\vec{H}$ is the magnetic field strength, $\vec{B}$ the magnetic flux density and $\vec{J}$ the electric current density. Further, we assume the affine material law $\vec{B} = \mu\vec{H} + \mu_0\vec{M}$, where $\vec{M}$ is the remanent magnetization of the permanent magnet \cite{Seibel:Moree_2022aa}.
The relative permeability $\mu_r = \tfrac{\mu}{\mu_0}$ of the different motor components -- as well as their color coding -- can be retrieved from
Fig.~\ref{Seibel:fig:PMSMandCD}. Completing, the magnetic vector potential $\vec{A}$ implicitly fulfilling the second equation of \eqref{Seibel:eq:MXWdivB0} is introduced as $\vec{B} = \seibelCurl \vec{A}$.

Assuming that electric current density and vector potential point in $z$-direction only, we can reduce the problem to two dimensions \cite{Seibel:Salon_1995aa}. The resulting equations read\vspace{-1em}

\begin{align}
	\mu\vec{h} - \seibelGrad a_z &= -\mu_0\vec{m},\label{Seibel:eq:MPP}\\
	-\seibelDiv \vec{h} &= j_z,\nonumber
\end{align}
with the $z$-component of the magnetic vector potential $a_z$, the rotated magnetic field strength $\vec{h}=(-H_y,H_x)^\top$, the permeability $\mu$, the vacuum permeability $\mu_0$, the rotated magnetization $\vec{m}=(-M_y,M_x)^\top$ and the $z$-component of the electric current density~$j_z$. 
Note, first, that we used the identity $\seibelCurlSD\vec{v} = \ \partial_x v_y - \partial_y v_x\ = -\seibelDiv \vec{v}^\perp,$ where $\vec{v}^\perp~=~(-v_y,v_x)^\top$. Second, in contrast to the formulation as a single second order PDE (e.g. (3.8) \cite{Seibel:Salon_1995aa}), we do not apply any differential operator to the magnetization $\vec{m}$. This is beneficial as $\vec{m}$ is not necessarily continuous and, hence, we circumvent the appearance of distributional derivatives.

\subsection{Domain Decomposition and Boundary Conditions}\label{Seibel:SubSec:DDBC}
The computational domain is split by the interface $\Gamma$, that is located in the middle of the air gap, into two disjoint subdomains $\Omega_\text{R}$ and $\Omega_\text{S}$ for the rotor and stator, respectively. The outer boundary of the electric machine and, hence, the stator is denoted by $\Sigma_\text{S}$ while the boundary between rotor and shaft is denoted by $\Sigma_\text{R}$. The boundaries towards neighboring poles are represented by $\Sigma^-$ and $\Sigma^+$. See Fig.~\ref{Seibel:fig:CompDom} for a sketch of the geometry.

\noindent
On the interface $\Gamma$ we choose boundary conditions respecting the tangential continuity of the magnetic field strength. Note, that $\vec{h}$ is rotated by \SI{90}{\degree} and, thus, being continuous in the normal but not tangential direction. On the outer boundaries $\Sigma_\text{S}$ and $\Sigma_\text{R}$, the magnetic vector potential $a_z$ fulfills homogeneous Dirichlet boundary conditions, while $a_z$ and $\vec{h}\cdot\vec{n}$ are anti-periodic on $\Sigma^-$ and $\Sigma^+$:
\hspace{-.5em}
\begin{subequations}
	\begin{alignat*}{2}
		a_z^+ &= a_z^- &&\quad\text{on } \Gamma,\\
		\vec{h}^+\cdot\textbf{n}_\Gamma &= \vec{h}^-\cdot\textbf{n}_\Gamma &&\quad\text{on } \Gamma,\\
		a_z &= 0 &&\quad\text{on }\Sigma_\text{S}\cup\Sigma_\text{R},\\
		a_z\left(\tfrac\pi3,r\right) &= -a_z\left(0,r\right) &&\quad\forall (0,r)\in\Sigma^-\\
		\vec{h}\cdot\vec{n}_{\Sigma}^+\left(\tfrac\pi3,r\right) &= \vec{h}\cdot\vec{n}_\Sigma^-\left(0,r\right) &&\quad\forall (0,r)\in\Sigma^-.%
	\end{alignat*}
\end{subequations}
By $a_z^\pm$ we denote the left- and right-handed traces of $a_z$ on $\Gamma$, $\Sigma^+$ and $\Sigma^-$, by $\vec{n}_\Sigma^-$ and $\vec{n}_{\Sigma}^+$ the outward normal vectors on $\Sigma^-$ and $\Sigma^+$, respectively. Note, that as we consider $n_\mathrm{seg}=6$ segments, the motor setup is anti-periodic with an angle of $\tfrac{2\pi}{n_\mathrm{seg}}=\tfrac{\pi}{3}$.

\section{Hybrid Mixed Domain Decomposition}\label{Seibel:sec:HMDD}
The most relevant part in the field of domain decomposition is the (re)coupling of local solutions to obtain a global solution. In this work, we use the \emph{hybrid mixed domain decomposition} (HMDD) method introduced in \cite{Seibel:Schmidt_2026aa}. The formulation can be transferred to the present magnetostatic rotor-stator problem without modification. The advantage of this approach is the reduced number of globally coupled degrees of freedom and the absence of problems at cross sections of subdomains as they arise in other domain decomposition methods such as Finite Element Tearing and Interconnecting (FETI) (see~e.g.~\cite{Seibel:Farhat_2001aa}).

\subsection{HMDD Method}
Inspired by \emph{hybridized discontinuous Galerkin} (HDG) methods (see e.g. \cite{Seibel:Cockburn_2009aa}) we introduce a hybrid variable $\lambda$, approximating the trace of $a_z$ on the interface $\Gamma$, to couple the subdomain problems and a penalization parameter $\tau$ is introduced to control the continuity of the finite element solution. Using the mixed problem \eqref{Seibel:eq:MPP} in combination with the boundary conditions discussed in Sec.~\ref{Seibel:SubSec:DDBC}, we obtain the following formulation:\\
\\
Seek $(\vec{h}_h,a_{z,h},\lambda_h)\in W_h\times Q_h\times M_h$ such that
\begin{align}
	\int_{\seibelOmG}\mu \vec{h}_h\cdot\vec{h}_h^\prime + a_{z,h} \seibelDiv\vec{h}_h^\prime \,\mathrm{d}\vec{x} + \int_{\Gamma_\Sigma} \lambda_h \seibelJump{\vec{h}_h^\prime\cdot\vec{n}}\,\mathrm{d}\sigma\nonumber
	&= -\mu_0\int_{\seibelOmG}\vec{m}\cdot\vec{h}_h^\prime\,\mathrm{d}\vec{x},\\
	-\int_{\seibelOmG} \seibelDiv\vec{h}_h a_{z,h}^\prime \,\mathrm{d}\vec{x} + \sum_\pm \int_{\Gamma_\Sigma}\tau(a_{z,h}^\pm-\lambda_h) a_{z,h}^{\prime\pm} \,\mathrm{d}\sigma%
	&= \int_{\seibelOmG} j_z a_{z,h}^\prime \,\mathrm{d}\vec{x}
	,\label{Seibel:eq:FEMFormulation}\\
	-\int_{\Gamma_\Sigma}\seibelJump{\vec{h}_h\cdot\vec{n}}\lambda_h^\prime\,\mathrm{d}\sigma - \sum_\pm \int_{\Gamma_\Sigma}\tau(a_{z,h}^\pm-\lambda_h)\lambda_h^\prime\,\mathrm{d}\sigma &= 0,\nonumber
\end{align}
for all $(\vec{h}_h^\prime,a_{z,h}^\prime,\lambda_h^\prime)\in W_h\times Q_h\times M_h$.\\
\\
\noindent
Here, $\seibelOmG = \Omega_\text{S}\cup\Omega_\text{R}$, $\Gamma_\Sigma = \Gamma\cup\Sigma^+\cup\Sigma^-$, and $\seibelJump{\vec{h}_h\cdot\vec{n}}$ and $\seibelJump{\vec{h}_h^\prime\cdot\vec{n}}$ denote the jump of $\vec{h}_h$ and $\vec{h}_h^\prime$ across $\Gamma$ or $\Sigma^\pm$, respectively. Formulation \eqref{Seibel:eq:FEMFormulation} fits into the HMDD framework \cite{Seibel:Schmidt_2026aa}, hence, we can transfer the properties of the HMDD method to \eqref{Seibel:eq:FEMFormulation}.

The space $W_h$ for the (rotated) magnetic field strength is chosen to be the Raviart-Thomas space of order $q$ on $\seibelOmG$, where discontinuities across $\Gamma$ are permitted. The spaces $Q_h$ and $M_h$ are discontinuous, piecewise polynomial spaces of order $q$ on $\seibelOmG$ and $\Gamma$, respectively.

This yields
\begin{align}
	W_h &\coloneq \{\vec{w}_h\in H(\seibelDiv,\seibelOmG),\ \vec{w}_h\circ\Phi_P\in\mathcal{RT}^q\ \forall K\in \mathcal{T}_h\},\\
	Q_h &\coloneq \{v_h|_K\circ\Phi_V \in \mathcal{P}^q\otimes\mathcal{P}^q\ \forall K \in \mathcal{T}_h\},\\
	M_h &\coloneq \{\nu_h|_F\circ\Phi_F\in\mathcal{P}^q\ \forall F\in\mathcal{F}_h\},
\end{align}
\noindent
where $\Phi_P$ is the Piola transform and $\Phi_V$ and $\Phi_F$ are area or length conserving transforms.

\subsection{Well-Posedness and Error Estimates}
In \cite{Seibel:Schmidt_2026aa} we were able to show that the HMDD method is well-posed, both in the continuous as well as in the discrete setting including finite elements:

For any right-hand side $\vec{f} \in L^2(\seibelOmG)^d$ there exist unique solutions $(\vec{h}_h, a_{z,h}, \lambda_h) \in W_h \times Q_h \times M_h$ of~\eqref{Seibel:eq:FEMFormulation} where it holds with a constant $C$ independent of $\tau$ that
\begin{subequations}
	\label{Seibel:eq:HDD:FEM:Wellposedness}
	\begin{align}
		\big\|\vec{h}_h\big\|_{H(\seibelDiv,\Omega)}
		+ \big\|a_{z,h}\big\|_{L^2(\seibelOmG)}
		+ \big\|\lambda_h\big \|_{L^2(\Gamma)}
		&\leq C \big\|f \big\|_{L^2(\seibelOmG)}, \\
		\big\|\seibelJump{\vec{h}_h\cdot\vec{n}}\big\|_{L^2(\Gamma)} &\leq C \sqrt{\tau} \|f\|_{L^2(\seibelOmG)},\label{Seibel:eq:HDD:FEM:NormEstqn}\\
		\sqrt{\tau}%
		\left(%
		\big\|\seibelJump{a_{z,h}}\big\|_{L^2(\Gamma)} + \big\| \left\{a_{z,h}\right\} - \mu_h\big\|_{L^2(\Gamma)}
		\right)%
		&\leq C \|f\|_{L^2(\seibelOmG)}\label{Seibel:eq:HDD:FEM:NormEstjumpu},\\
		\sum_\pm \tfrac{1}{\sqrt{1+\tau}} \big\| \sqrt{\tau} a_{z,h}^\pm \big\|_{L^2(\Gamma)} &\leq C \big\|f \big\|_{L^2(\seibelOmG)}.
	\end{align}
\end{subequations}
Note, that the estimates partially depend on the stabilization parameter $\tau$ and that in our setting $\vec{f}=(\vec{m},j_z,0)^\top$. From \eqref{Seibel:eq:HDD:FEM:NormEstqn} we deduce that the continuity of $\vec{h}_h\cdot\vec{n}$ is increasingly enforced for decreasing $\tau$, while \eqref{Seibel:eq:HDD:FEM:NormEstjumpu} shows the continuity of $a_{z,h}$ being increasingly enforced for increasing $\tau$.

We, further, conjectured error estimates in the case of sufficiently smooth solutions~\cite{Seibel:Schmidt_2026aa}, which we can assume to hold for \eqref{Seibel:eq:FEMFormulation} as well:
	\begin{subequations}
	\begin{align}
		\| a_{z} - a_{z,h} \|_{L^2(\seibelOmG)} + \| \lambda - \lambda_h \|_{L^2(\Gamma)} &\leq
		C_1 h^{q+1},
		\label{Seibel:eq:eoc:1}\\
		\| \vec{h} - \vec{h}_h \|_{L^2(\seibelOmG)} &\leq
		C_1 h^{q+1} + \tfrac{C_2 h\tau}{C_3 + h\tau} h^{q+\frac12},
		\label{Seibel:eq:eoc:2}
	\end{align}
	\label{Seibel:eq:conj}
\end{subequations}\\
\noindent
and we expect~\eqref{Seibel:eq:eoc:2} to hold for the discretization error in the magnetic flux density $\vec{b}_h~=~\mu \vec{h}_h$. %
We found that the convergence rates depend on the product of the mesh width and the stabilization parameter $\tau$. For a more extensive discussion of this behavior we refer the reader to \cite{Seibel:Schmidt_2026aa}.

\begin{figure}[b]
	\centering
	\includegraphics[height=4.5cm]{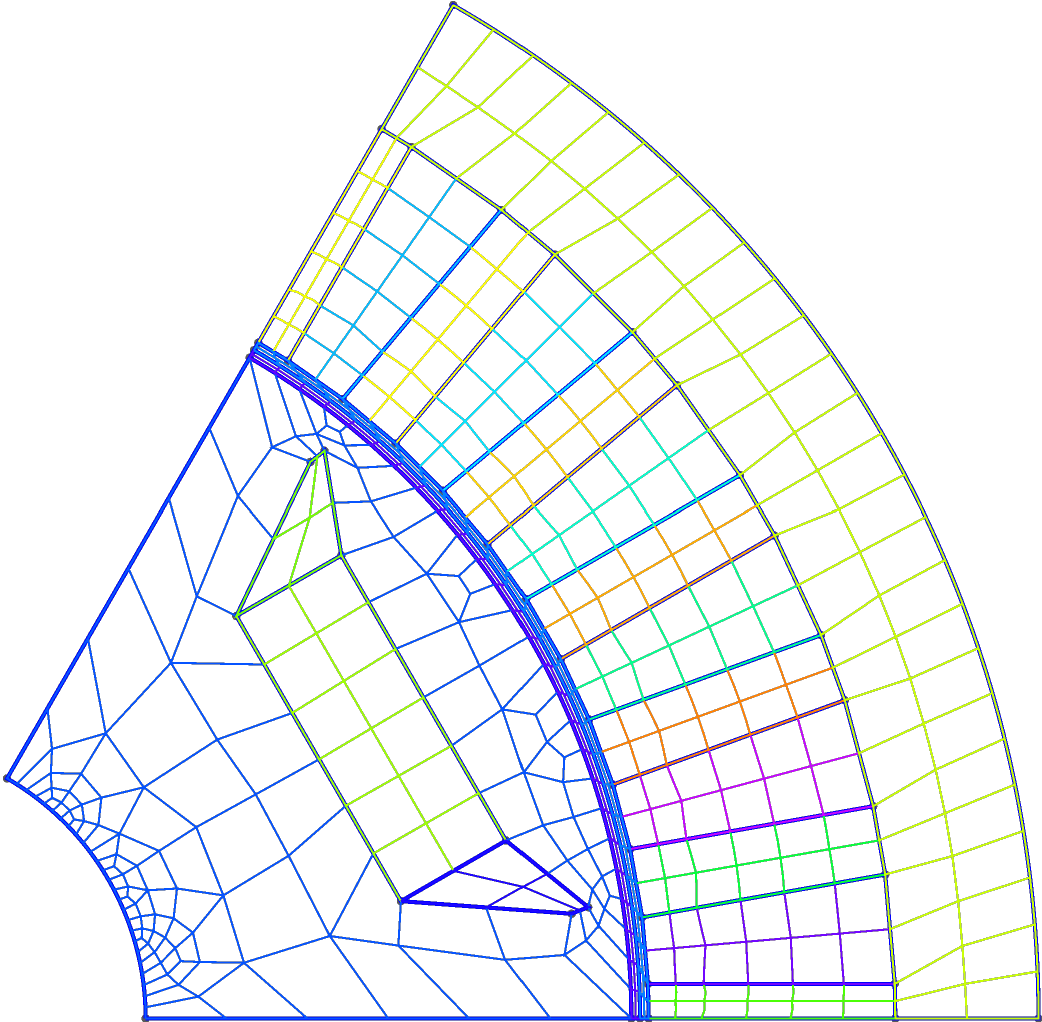}
	\caption{Sketch of the initial quadrilateral mesh used for the experiment.}
    \label{Seibel:fig:LOMesh}
\end{figure}

\section{Numerical Experiments}
The experiments are performed using the numerical \texttt{C++}-library \texttt{Concepts~2}~\cite{Seibel:ConceptsDevelopmentTeam_2026aa}, which provides interfaces to various direct solvers solvers -- for this work we used \texttt{SuperLU}. It supports being run in parallel, and is suitable for high-order and $hp$-adaptive FEM \cite{Seibel:Schmidt_2009ab}. In particular, \texttt{Concepts} supports quadrilateral mesh elements with circularly curved boundaries, which is beneficial for exact geometry representation. It can read \texttt{gmsh} mesh files and we used the interface to \texttt{MATLAB} for graphical representation. The initial mesh as depicted in Fig.~\ref{Seibel:fig:LOMesh}.

\noindent
We perform an experiment that serves as a proof of concept reinforcing the applicability of HMDD for the simulation of a PMSM. Therefore, we assume that adjacent slots pairwise share a phase in the current density. The phase between two neighboring pairs is shifted by \SI{120}{\degree}. We, further, assume the current density to be $\left|j_z\right|=\SI{5}{\ampere\per\milli\meter\squared}$ and the remanence being $\left|\mu_0\vec{m}\right|=\SI{1}{\tesla}$. Additionally, we set the stabilization parameter to $\tau=\num{1.0}$ and consider FE spaces of order $q=1$.

\begin{figure}[b]
	\begin{subfigure}{.45\textwidth}
		\includegraphics[height=4.5cm]{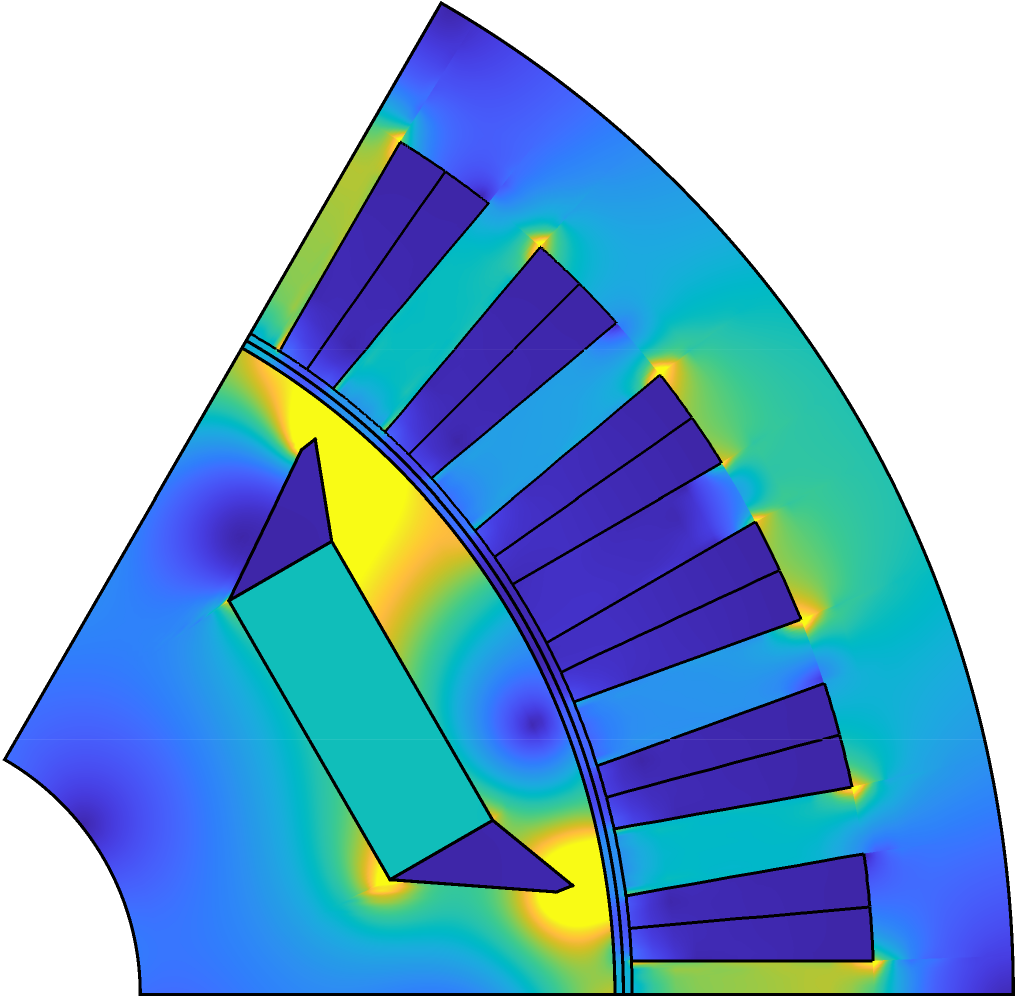}
	\end{subfigure}
	\begin{subfigure}{.45\textwidth}
		\includegraphics[height=4.5cm]{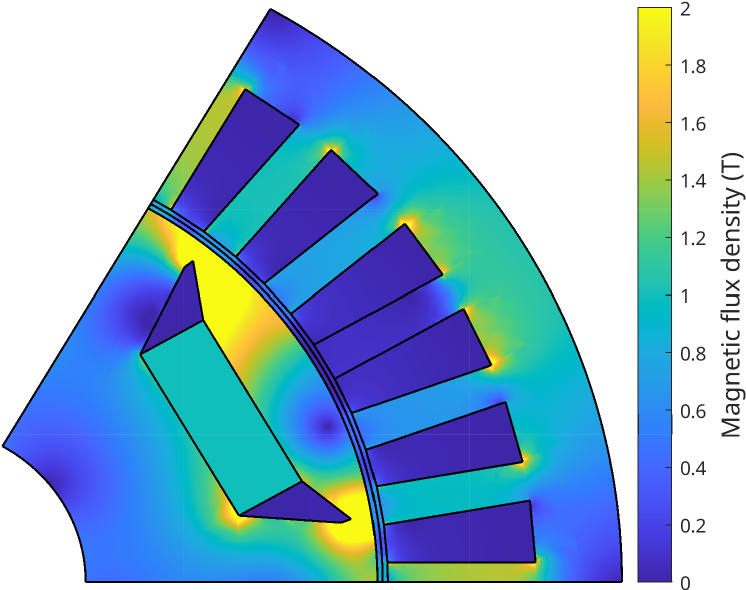}
	\end{subfigure}
	\caption{Resulting absolute value $|\vec{b}|$ of the magnetic flux density $\vec{b}$ using the in-house IGA code (left) and the HMDD method implemented with \texttt{Concepts 2} (right).}
    \label{Seibel:fig:bField}
\end{figure}

\begin{figure}[t]
	\begin{subfigure}{.45\textwidth}
		\includegraphics[height=4.5cm]{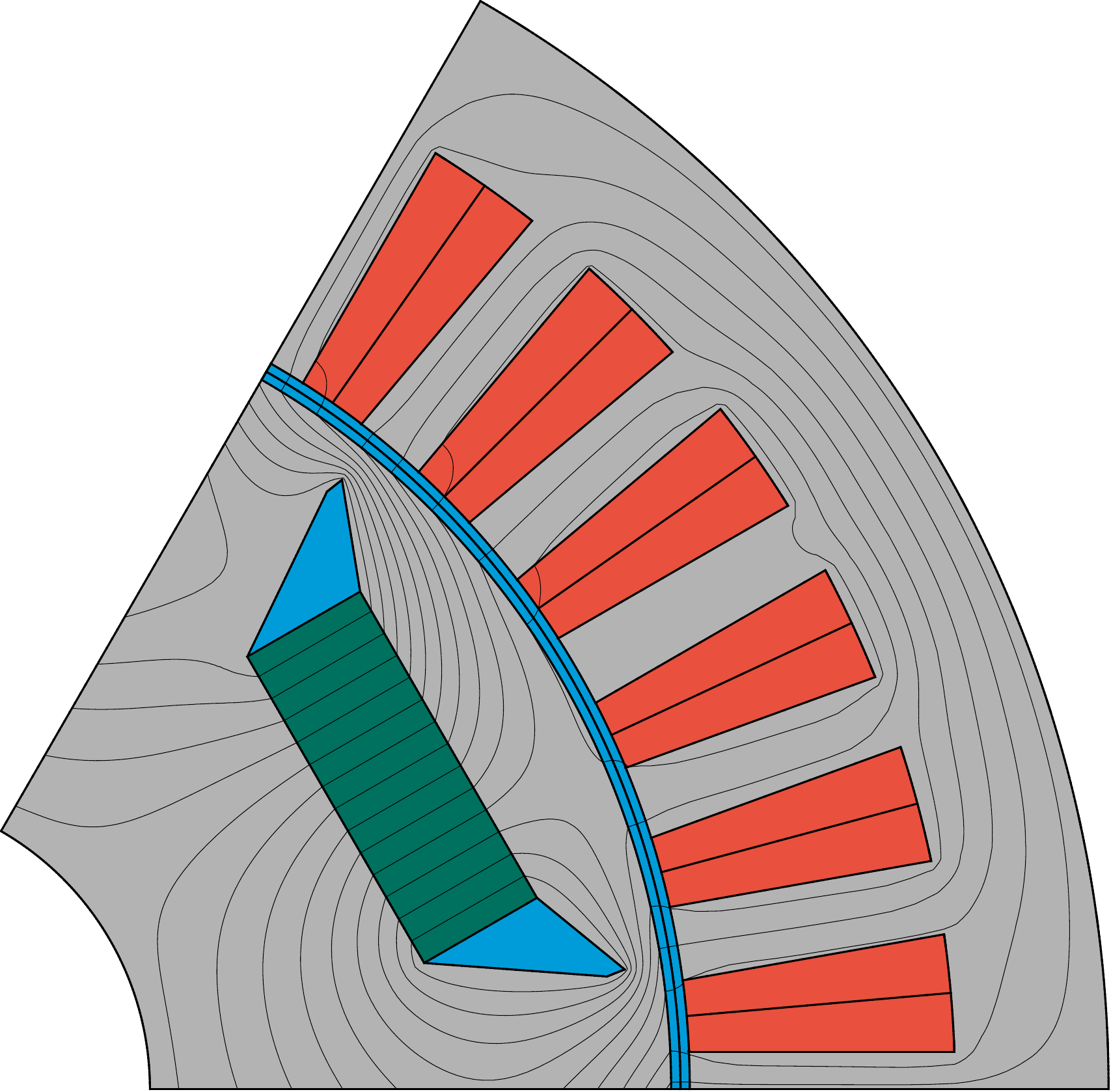}
	\end{subfigure}
	\begin{subfigure}{.45\textwidth}
		\includegraphics[height=4.5cm]{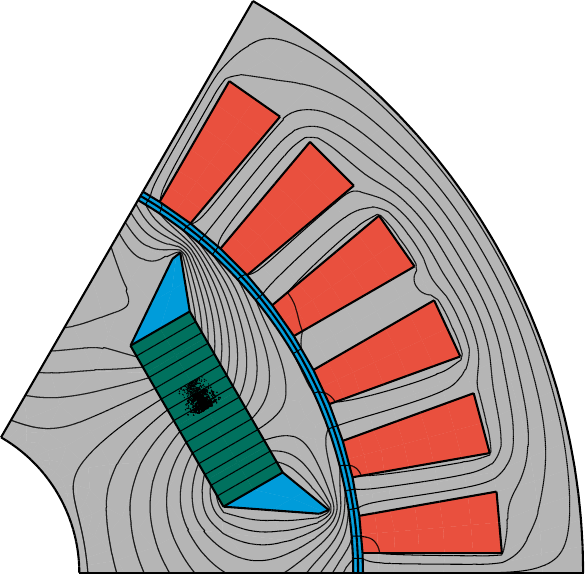}
	\end{subfigure}
	\caption{Resulting magnetic potential lines using the in-house IGA code (left) and the HMDD method implemented with \texttt{Concepts 2} (right).}
    \label{Seibel:fig:PotLines}
\end{figure}

The results are compared with those obtained by an in-house \emph{iso-geometric analysis} (IGA) code \cite{Seibel:Wiesheu_2023ag} to assess the approach's accuracy.
First, we consider the magnitude of the (rotated) magnetic flux density $|\vec{b}_h|$. On the left-hand side of Fig.\ref{Seibel:fig:bField} we see the results obtained by the in-house IGA code while the right-hand subfigure shows the result of the HMDD method implemented in \texttt{Concepts}.
In Tab.~\ref{Seibel:tab:bh_norm} 6.5~Tthe $L^2$-norm of $\vec{b}_h$ is shown for the HMDD method on the original mesh and an once refined mesh.

	\begin{table}[]
		\caption{Mesh width $h$, number of degrees of freedom (ndof) and $L^2$-norm of the magnetic flux density computed with HMDD on the original mesh and a once refined mesh of different mesh width $h$ and number of degrees of freedoms (ndof).}
		\label{Seibel:tab:bh_norm}
		\centering
		\begin{tabular}{c@{\hspace{2em}}r@{\hspace{2em}}c}
			$h$ & ndof & $\|\vec{b}_h\|_{L^2(\Omega)}$ \\ \hline
			\num{0.00972}  & \num{3852}  & \num{0.042424} \\
			\num{0.00529}  & \num{15128} & \num{0.042301}   \\ \hline
		\end{tabular}
	\end{table}
Second, we visualize the magnetic field lines given by the iso-potential lines of $a_{z,h}$ in Fig.~\ref{Seibel:fig:PotLines}.

The difference in the flux density measured in the $L^2$ norm is about \SI{16}{\percent} where it is for the magnetic potential about \SI{1.5}{\percent}. We observe that the $L^2$-norm of $\vec{b}_h$ changes for the HMDD solution from the original mesh to an once refined mesh by only \SI{0.2}{\percent}, i.e., the HMDD solution is closer to the exact solution than the IGA solution.
On the material corners of iron with other material singularities appear where the magnetic flux density becomes unbounded. %
We observe that the highest magnitude of $\vec{b}_h$ is about \SI{6.5}{\tesla} for the IGA solution where it is around \SI{13.2}{\tesla} for the HMDD solution and \SI{17.5}{\tesla} on a once refined mesh.
Here, the the mixed formulation in which the magenetic field is an unkown and the lower regularity of the FE basis functions shows an higher accuracy close to material corners.

\section{Conclusion \& Outlook}
We have derived a finite element method from a mixed magnetostatics problem in two dimensions that fits into the HMDD framework. The properties of the HMDD method are naturally transported to the considered example which then was implemented in \texttt{Concepts}. In an academic example, serving as proof of concept, we found good agreement with an in-house IGA code.\\
In a next step, we aim to advance to magnetoquasistatics with rotational motion by using a time-stepping procedure.\\
Regarding motion, we have the reasoned hope that the presented formulation is well-suited for mortaring as this is inherently supported by HDG methods for $0~<~\tau~<~\infty$~\cite{Seibel:Cockburn_2009aa}.

\section*{Acknowledgements}
This work is supported by the Graduate School CE within the Profile Topic Computational Engineering at the Technical University of Darmstadt. The authors thank Michael Wiesheu, funded by the Collaborative Research Centre – TRR361/F90: CREATOR, for the provision of Fig.~\ref{Seibel:fig:PMSM} and the comparison results in Fig.~\ref{Seibel:fig:bField} and Fig.~\ref{Seibel:fig:PotLines}. The first author thanks Christian Bergfried, also funded by CREATOR, for fruitful discussions and helpful insights on this topic.
\bibliographystyle{ieeetr}

\end{document}